\documentclass[twocolumn,floatfix,aps,superscriptaddress,showpacs]{revtex4}
\usepackage{graphicx}

\newcommand{\bc}{\begin{center}}
\newcommand{\ec}{\end{center}}
\newcommand{\nin}{\noindent}

\newcommand{\be}{\begin{equation}}
\newcommand{\ee}{\end{equation}}
\begin{document}
\title{Long-range energy-level interaction in small metallic particles}
\author{R. A. Jalabert}
\affiliation{C.E.A., Service de Physique de l'Etat Condens\'e,
Centre d'\'Etudes de Saclay,
91191 Gif-sur-Yvette C\'edex, France}
\author{J.-L. Pichard}
\affiliation{C.E.A., Service de Physique de l'Etat Condens\'e,
Centre d'\'Etudes de Saclay,
91191 Gif-sur-Yvette C\'edex, France}
\author{C. W. J. Beenakker}
\affiliation{Instituut-Lorentz, University of Leiden, P.~O.~Box 9506, 2300 RA Leiden,
The Netherlands}
\date{February 1993}
\begin{abstract}
We consider the energy level statistics of non-interacting electrons
which diffuse in a $d$-dimensional disordered metallic
conductor of characteristic
Thouless energy $E_c$. We assume that the level distribution can be written
as the Gibbs distribution of a classical one-dimensional gas of
fictitious particles with a pairwise additive interaction potential
$f(\varepsilon)$.
We show that the interaction which is consistent with the
known
correlation  function of pairs of energy levels
is a logarithmic repulsion for level separations
$\varepsilon < E_c$, in agreement with Random Matrix Theory.
When $\varepsilon > E_c$, $f(\varepsilon)$ vanishes as
a power law in $\varepsilon / E_c$ with exponents
$-\frac{1}{2}$,$-2$, and $-\frac{3}{2}$ for $d=1$, $2$, and $3$, respectively.
While for $d=1,2$ the energy-level interaction is always
repulsive, in three dimensions there is long-range level
attraction after the short-range
logarithmic repulsion.
\end{abstract}
\pacs{02.50.+s, 05.45.+b, 05.60.+w, 72.15.Rn}
\maketitle

   A statistical description of the hamiltonian {\bf H}
of a complex
system is provided by random matrix theory. A key feature in
this theory is the {\it spectral rigidity} of
the energy levels: their distribution $P(E_{1},E_{2},...E_{N})$
formally coincides with the Gibbs distribution of the
positions of a one-dimensional gas of $N$ classical particles with a
{\it repulsive logarithmic interaction},

\FL
\begin{equation}
P(\{E_{n}\}) = Z^{-1} \exp{[-\beta {\cal H}(\{E_{n}\})]} \ ,
\label{eq:P}
\end{equation}

\begin{equation}
{\cal H}(\{E_{n}\})= - \sum_{i<j}\ln{|E_{i}-E_{j}|}+ \sum_{i} V(E_{i})
\ .
\label{eq:H}
\end{equation}

\nin Here $Z$ is a normalization constant, and $V(E)$ is a confining
potential.  The parameter $\beta$, playing the role of an inverse temperature,
depends on the symmetry class of the ensemble
 of random hamiltonians \cite{mehta}.
 A method which yields such a distribution
consists in assigning to the hamiltonian {\bf H}
a probability distribution
of maximum information entropy given a spectral constraint  \cite{balian}.
For instance, a constraint on the expectation value of $\sum_i E_i^2$
yields the gaussian ensembles, where $V(E) \propto E^2$.
Other ensembles, characterized by different $V(E)$, result from
from other spectral constraints (e.g. the averaged level density).
All these classical ensembles of random matrices have in common an
absence of eigenvalue--eigenvector correlations and a logarithmic
repulsion between pairs of eigenvalues.

The use of this theory for the study of electronic properties
of small metallic particles was introduced by Gorkov and
Eliashberg \cite{gorkov}.
Theoretical support for the logarithmic repulsion of energy levels came
with the work of Efetov \cite{efetov}. Assuming bulk diffusion
of the electrons by elastic scatterers and using a supersymmetric
formalism, he obtained for the
spectrum of small metallic particles the same correlation function as in
classical sets of random matrices. Subsequently, addressing the connection
between universal conductance fluctuations \cite{RevWebb} and the
universal properties of random matrices,
Al'tshuler and Shklovski\u{\i} \cite{AS} (A\&S) showed that
for energy separations $|E-E'|$ greater than the Thouless energy $E_{c}$,
the correlation function deviates from classical
random matrix theory. The Thouless energy $E_{c} =D\hbar/L^2$ is inversely
proportional to the time $t_{\rm erg}$ it takes an electron to diffuse
(with diffusion coefficient $D$)
across a particle of size $L$. The results of the diagrammatic perturbation
theory of A\&S were recently rederived by Argaman, Imry, and Smilansky,
using a more intuitive semiclassical method \cite{AIS}.

 One would not expect that the logarithmic level repulsion in Eq.
(\ref{eq:H}) holds for levels which are separated by more than
$E_{c}$. What is then the long-range energy-level interaction in small
metallic particles? This is the question addressed in this paper. In a
sense, this is an inverse problem in statistical mechanics: given the
pair correlation function, what is the interaction potential? Our
analysis applies a recently developed functional-derivative technique to
compute correlation functions in random-matrix ensembles with an
arbitrary two-body interaction potential \cite{Been}. The restriction to
two-body (i.e. pairwise additive) interaction is our single assumption.
We find that $E_c$ characterizes a cross-over between a short-range
logarithmic repulsion and a novel long-range part which
decays as a power law, with a dimensionality-dependent exponent.
The interaction remains repulsive for dimensions 1 and 2, but exhibits a
long-range {\it attractive} part after a short-range repulsion in 3 dimensions.

The starting point of our analysis is the Gibbs distribution
(\ref{eq:P}) with an arbitrary two-body interaction $f(|E-E'|)$ in
the fictitious ``hamiltonian" ${\cal H}$,

\begin{equation}
{\cal H}(\{E_{n}\})= \sum_{i<j}f(|E_{i}-E_{j}|)+ \sum_{i} V(E_{i}) \ .
\label{eq:HG}
\end{equation}
The mean eigenvalue density $\langle \rho(E)\rangle$ \
($\sum_{n} \delta(E-E_{n})$ (where $\langle\ldots\rangle$ denotes an
average with weight $P(\{E_{n}\})$) is related to $V(E)$ and $f(E-E')$
by an integral equation, valid\cite{Dy} to the leading order of a
$1/N$ expansion:
\begin{equation}
V(E) = - \int_{-\infty}^{\infty} dE' \langle \rho(E')\rangle
f(|E-E'|) + c \ .
\label{eq:MF}
\end{equation}
\nin The constant $c$ is to be determined from the normalization condition
$\int dE \langle \rho(E) \rangle =N$. Eq. (\ref{eq:MF}) has the intuitive
``mean-field" interpretation (originally due to Wigner), that the
``charge density" $\langle \rho \rangle$ adjusts itself to the
``external potential" $V$ in such a way that the total force on any
charge $E$ vanishes. Dyson \cite{Dy} has evaluated the first correction
to Eq. (\ref{eq:MF}), which is smaller by a factor $N^{-1}\ln{N}$.

The density-density correlation function defined by
\begin{equation}
K_{2}(E,E') = \langle \rho(E)\rangle \langle\rho(E')\rangle -
\langle \rho(E) \rho(E') \rangle \ ,
\label{eq:K}
\end{equation}
\nin can be expressed as a functional derivative \cite{Been},
\begin{equation}
K_{2}(E,E') = \frac{1}{\beta} \frac{\delta \langle \rho(E)\rangle}
{\delta V(E')} \ .
\label{eq:FD}
\end{equation}

Eq. (\ref{eq:FD}) is an exact consequence of Eqs. (\ref{eq:P}) and
(\ref{eq:HG}). Physically, it means that correlations
between $E$ and $E'$ are important when a modification of
the potential at $E'$ has a substantial impact on  the mean
density at $E$. Combining Eqs. (\ref{eq:MF}) and (\ref{eq:FD}),
one can see that $K_{2}(E,E') \equiv K_{2}(|E-E'|)$ is
translationally invariant and independent of
the confining potential $V(E)$, depending on the two-body interaction
$f(\varepsilon)$ only. This property is at the heart of universality in
random-matrix theory \cite{Been,Dy}.

By Fourier transforming the convolution (\ref{eq:MF}),
the time-dependent two-level form factor

\begin{equation}
K_{2}(t) = \int_{-\infty}^{\infty}
d\varepsilon K_{2}(\varepsilon) \exp{\left(-\frac{i\varepsilon t}
{\hbar}\right)}
\label{eq:ftk2}
\end{equation}

\nin can be written as

\begin{equation}
K_{2}(t) = \frac{1}{\beta} \frac{\delta \langle \rho(t)\rangle}
{\delta V(t)} = - \frac{1}{\beta f(t)} \ .
\label{eq:oof}
\end{equation}

\nin This relationship gives us the prescription for
obtaining the eigenvalue interaction $f(\varepsilon)$
from the density-density correlation function $K_2(\varepsilon)$.

For disordered systems in the weak-scattering limit, A\&S have shown
by perturbation theory that the correlation function is
given by

\begin{equation}
K_{2}(\varepsilon) = \frac{s^2}{\beta \pi^2}
Re \sum_{\{n_{\mu}\}} (\varepsilon + i\hbar D q^2 + i\gamma)^{-2} \ ,
\label{eq:AS}
\end{equation}

\nin for energies $\varepsilon$ large compared to the level spacing
$\Delta$, and small compared with the energy scale $\hbar / \tau_e$,
associated with the elastic scattering time $\tau_e$.
The factor $s=2$ accounts for the spin degeneracy
of each level, $\gamma$ is a small energy cutoff (to account for
inelastic scattering) and the parameter $\beta$ equals 1 (2) in the
presence (absence) of time-reversal symmetry ($\beta=4$ for
time-reversal symmetry with strong spin-orbit scattering).
The sum is over the eigenvalues of the diffusion equation
for the sample, assumed to be a $d$-dimensional parallelepiped with sides
$L_{\mu}$ ($q^2=\pi^2 \sum_{\mu=1}^{d}(n_{\mu}/L_{\mu})^2$). In
what follows we put $s=1$ and $\gamma=0$, ignoring the
spin degeneracy and the small-energy cutoff. The Fourier transform
of Eq. (\ref{eq:AS}) is

\begin{equation}
K_{2}(t) = - \frac{|t|}{\beta \pi \hbar}
\sum_{\{n_{\mu}\}} \exp{\left(-D\pi^2|t|
\sum_{\mu=1}^{d}(n_{\mu}/L_{\mu})^2\right)} \ .
\label{eq:FTAS}
\end{equation}

The long- and short-range limits $K_{2}^{l}(t)$ and $K_{2}^{s}(t)$ of the
form factor (\ref{eq:FTAS}) can be obtained in closed form \cite{AIS}.
The cross-over time scale is the so called ergodic time $t_{\rm erg}=
L^2/D$, which it takes an electron to explore the whole available phase
space (for simplicity we work with a hypercube,
$L_{\mu}=L$ for all $\mu$). The ergodic time is related to the Thouless
energy by $E_{c}=\hbar /t_{\rm erg}$. For times $t \gg t_{\rm erg}$
the first term of Eq. (\ref{eq:FTAS}) dominates the sum,
while for $t \ll t_{\rm erg}$ one
can convert the sums over $n_{\mu}$ into gaussian integrals. The
resulting long- and short-time limits are \cite{AIS}

\begin{equation}
K_{2}^{l}(t) = - \frac{|t|}{\beta \pi \hbar} \ ,
\label{eq:K2l}
\end{equation}

\begin{equation}
K_{2}^{s}(t) =  - \frac{|t|}{\beta \pi \hbar}
\frac{L^d|t|^{-d/2}}{(4 \pi D)^{d/2}} \ .
\label{eq:K2s}
\end{equation}

For analytical work it is convenient to have an expression which
smoothly interpolates between these two asymptotic limits. We will
use the interpolation formula
\begin{equation}
K_{2}(t) \simeq K_{2}^{l}(t) + K_{2}^{s}(t) \ .
\label{eq:Inter}
\end{equation}
The approximation (\ref{eq:Inter}) differs from the full expression
(\ref{eq:FTAS}) for $t \simeq t_{\rm erg}$, but it is accurate for $t$
either much smaller or much larger then $t_{\rm erg}$. This is sufficient
for the purpose of obtaining the asymptotic behavior of the interaction
potential.

Combining Eqs. (\ref{eq:FD}) and (\ref{eq:Inter}), the interaction
potential $f_{d}(\varepsilon)$ for $d$ dimensions can be written as

\begin{equation}
f_{d}(\varepsilon) = \int_{0}^{\infty} dt \ \cos{(\alpha t)} \
\frac{t^{d/2}}{t(1+t^{d/2})} \ ,
\label{eq:fd}
\end{equation}

\nin where $\alpha =\varepsilon/(4\pi E_{c})$ is the dimensionless energy
variable. The integral (\ref{eq:fd}) can be evaluated numerically for
all $\alpha$, and analytically in the small- and large-$\alpha$ limits.
For $|\alpha| \ll 1$ we can approximate $\cos{\alpha t} \simeq 1$ and
cut the upper integration limit at $1/|\alpha|$, which readily yields the
short range universal logarithmic interaction
$f_{d}(\varepsilon) \simeq - \ln{|\alpha|}$. For
$|\alpha |\rightarrow \infty$ the high-frequency oscillations
of $\cos{\alpha t}$ average the integral to zero, in a way which
depends on the dimensionality. The easiest case is $d=2$,
where the integral can be evaluated in a closed form,

\begin{equation}
f_{2}(\varepsilon) = - \sin{|\alpha|} \ {\rm si}|\alpha| \ -
\cos{(\alpha)} \ {\rm ci}(\alpha) \  ,
\label{eq:f2}
\end{equation}

\nin which behaves  as $- \ln{|\alpha|} - {\cal C}$
for small $|\alpha|$ (${\cal C}$ is Euler's constant), and $1/\alpha^2$
is the dominant term of an asymptotic expansion of Eq. (\ref{eq:fd})
for $|\alpha| \gg 1$. We therefore recover
the short-range logarithmic repulsion and find that
the interaction remains repulsive in the whole energy range. For $d=3$, the
asymptotic limits of the interaction can be obtained by considering the
auxiliary function

\begin{equation}
h(\alpha) = \int_{0}^{\infty} dt \ \frac{\sin{\alpha t}}{t} \
\frac{t^{1/2}}{1+t^{3/2}} \ ,
\label{eq:haux}
\end{equation}

\nin which satisfies $h'(\alpha)=f_{3}(\alpha)$ as well as the
differential equation

\begin{equation}
h'(\alpha) = \frac{1}{\alpha} h(\alpha) -
\frac{3}{\alpha} \int_{0}^{\infty} du \ \frac{\sin{(\alpha u^2)}}
{(1+u^3)^2} \ .
\label{eq:diffeq}
\end{equation}

\nin The second term on the r.h.s. becomes $-1$ in the small-$\alpha$
limit, and $- 3\sqrt{\pi}|2\alpha|^{-3/2}$ in the large-$\alpha$ limit.
We thus obtain $f_{3}(\varepsilon) \simeq - \ln{|\alpha|}-{\cal C}$ for
$|\alpha| \ll 1$ and
$ - \sqrt{\pi}|2\alpha|^{-3/2}$ for $|\alpha| \gg 1$.
Therefore, in $d=3$, we have an {\it attractive} eigenvalue interaction for
large separarations. Using a similar
procedure for $d=1$ we obtain the same short-range logarithmic repulsion
$- \ln{|\alpha|}-{\cal C}$, which crosses over to an algebraic repulsion
$f_{1}(\varepsilon) \simeq \sqrt{\pi}|2\alpha|^{-1/2}$ for $\alpha \gg 1$.

In Fig. 1 we compare a numerical integration of Eq. (\ref{eq:fd}) with
the asymptotic expressions derived above, which we summarize:

\begin{equation}
f_{d}(\varepsilon) =  - \ln{\left|\frac{\varepsilon}{4\pi E_{c}}\right|}
-{\cal C} \hspace{2cm} {\rm if} \ |\varepsilon| \ll E_{c} \ ,
\label{eq:fl}
\end{equation}

\begin{equation}
\left. \begin{array}{l}
f_{1}(\varepsilon) =  \left(\frac{\pi}{2}\right)^{1/2}
\left|\frac{\varepsilon}{4\pi E_{c}}\right|^{-1/2} \\
f_{2}(\varepsilon) = \left(\frac{\varepsilon}{4\pi E_{c}}\right)^{-2} \\
f_{3}(\varepsilon) =  - \frac{1}{2} \left(\frac{\pi}{2}\right)^{1/2}
\left|\frac{\varepsilon}{4\pi E_{c}}\right|^{-3/2}
\end{array} \right\}
\ {\rm if} \ |\varepsilon| \gg E_{c} \ .
\label{eq:fs}
\end{equation}
Fig. 1 shows the crossover from the universal logarithmic short-range
repulsion into the novel long-range power-law regime (repulsive
for $d=1,2$ and attractive for $d=3$). We can see that for a given
$E_{c}$ the departure from the universal regime occurs earlier as we go
to lower dimensions.

In summary, we have calculated in the metallic regime the
dimensionality--dependent
long-range part of the energy-level interaction.
{}From this interaction, one could in principle calculate
{\it n}-point correlation functions for arbitrary {\it n}.
Our method is based on a general relation \cite{Been} between the
density-density correlation function and this interaction.
We use it in a particular case where this correlation function
is known from diagrammatic perturbation theory
\cite{AS} (or an equivalent semiclassical theory \cite{AIS}). The
validity of our results is restricted
to the validity of these perturbative or semiclassical approaches:
$\varepsilon \gg \Delta$, and $\varepsilon \ll \hbar / \tau_e$.
Since Efetov has shown that random matrix theory remains valid
for  $\varepsilon \ll \Delta$, the universal logarithmic repulsion
which we recover must be also valid for these small energy
separations, though either perturbation theory (Eq. {\ref{eq:AS}}),
or our method based on an asymptotic large-N approximation
miss fine structure on the scale of $\Delta$.

The condition $\varepsilon \ll  \hbar / \tau_e$
limits the non-universal algebraic decay which we find
for $\varepsilon >E_c\equiv\hbar/t_{\rm erg}$.
These non-universal interactions
result from the non-ergodic electron dynamics for $ t < t_{\rm erg}$.
The non-ergodic dynamics in our problem is unbounded diffusion in
$d$ dimensions.
For times smaller than $\tau_{e}$ the electron motion
is ballistic, a behaviour which is not considered in our theory.

Chaotic billiards with ballistic motion
between the boundaries constitute another example for the applicability
of random-matrix theory. In the
semiclassical limit the energy level correlations are also
correctly described by the classical ensembles for energy
differences smaller than the
inverse period of the shortest periodic orbits of the system
\cite{BGS,Berr}. The essential difference with the small metallic particles
considered in this work is that these have a well-defined
statistical regime of diffusive motion between $\tau_{e}$
and $t_{\rm erg}$.

\begin{figure}[tb]
\centerline{\includegraphics[width=0.9\linewidth]{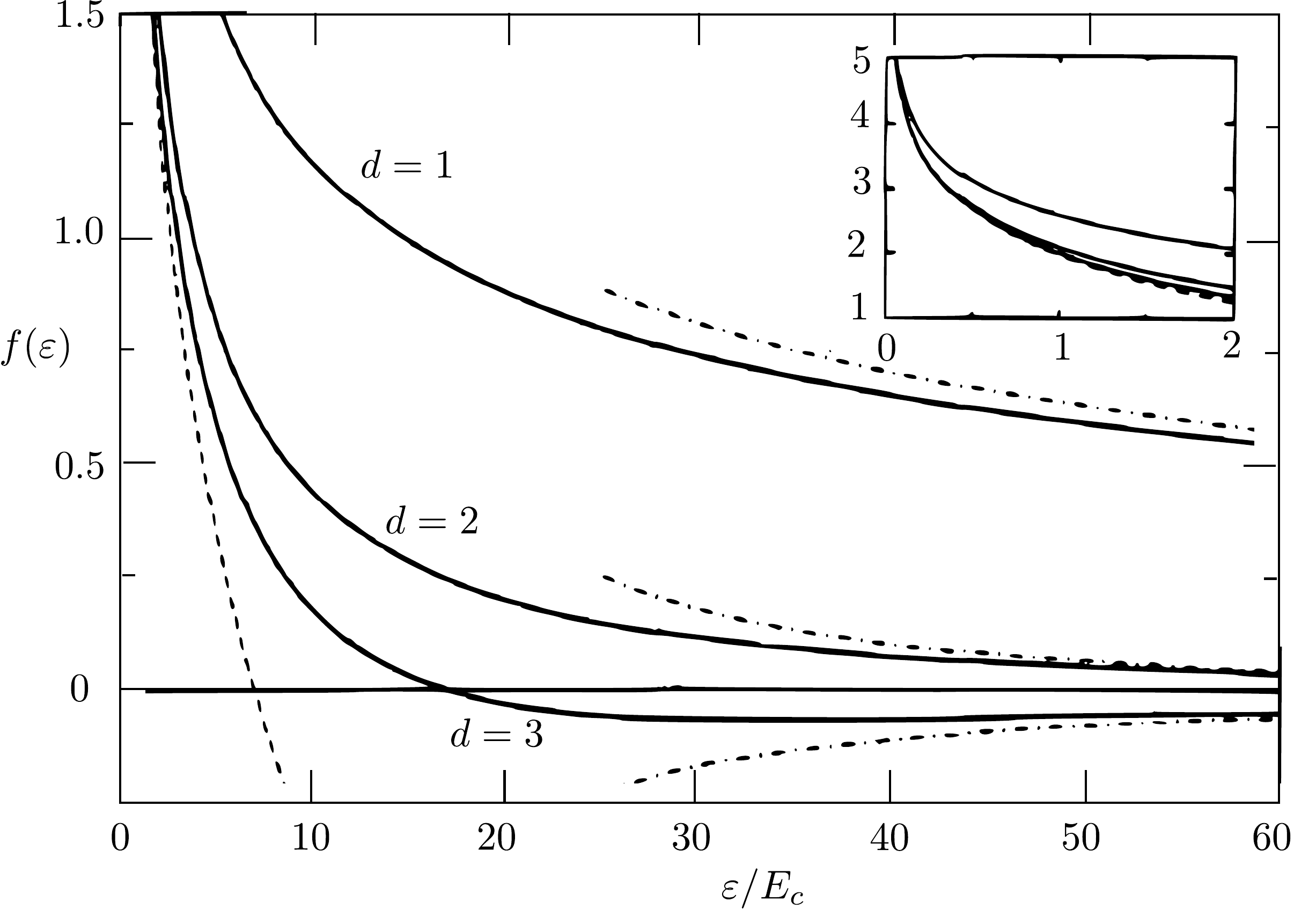}}
\caption{Interaction potential according to Eq. (\ref{eq:fd}) for various
spatial dimensions $d$ (solid), together with the
short-range logarithmic (dash) and long-range power law (dot-dash)
asymptotic forms described by Eqs. (\ref{eq:fl})
and (\ref{eq:fs}).
Inset: blow up of the departure from the short-range
logarithmic interaction.
\label{fig1}}
\end{figure}

It is interesting to consider the scale dependence of the
interaction potential. In $d=3$, the level spacing scales as
$\Delta \propto L^{-3}$ while $E_c \propto L^{-2}$, so that
$\Delta \ll E_{c}$ as $L \rightarrow \infty$. Therefore,
if we measure $|E-E'|$ in units of $\Delta$, the interaction $f(E-E')$
scales with $L$ towards the universal random matrix repulsion
for about $L$ nearest neighbour levels for three dimensional
conductors. However, the total number of levels
being proportional to $L^3$, the relation (\ref{eq:MF}) between the
average density $\langle \rho(E)\rangle$ and the confining
potential $V(E)$ still differs from the usual expression (i.e.
with a logarithmic interaction) in the thermodynamic limit.

Our analysis is restricted to metallic particles which are small
compared with the localization length.
In the thermodynamic limit, electrons are always localized  for
$d=1$ or 2 (except for $\beta=4$ in $d=2$ at low disorder).
Anderson localization occurs also in three dimensions for large
disorder. In these cases, our perturbative starting point Eq.
(\ref{eq:AS}) is no longer valid. In the presence
of eigenvector localization, $f(\varepsilon)$ probably
scales with the system size towards a delta function (uncorrelated levels
in the limit of strong localization). An interesting issue that
we postpone for
future studies is to see if the mobility edge is characterized
by some scale invariant interaction  $f_c(\varepsilon)$.

In this paper we have only considered the eigenvalue statistics.
The classical random matrix ensembles are invariant under canonical
(orthogonal, unitary or symplectic) transformations. The logarithmic
level repulsion is related to the maximum randomness of the
eigenvectors. In the localized regime, it is clear that the absence
of a logarithmic level repulsion has its physical origin in the
localization of the eigenvectors in different parts of the system.
An important issue would consists in
identifying what kind of modification of the eigenvector statistics
from Porter-Thomas distribution \cite{mehta} is behind the novel
non-universal part of the level interaction found in this work.

Finally, it is likely that studies similar to the one
presented here for the hamiltonian ensemble will be useful for the
ensemble of scattering or transfer matrices\cite{St}, and will
improve our understanding of quantum transport in disordered conductors.

This work was supported in part by EEC, Contract No. SCC--CT90--0020,
and by the Dutch Science Foundation NWO/FOM.

\end{document}